\documentclass[pre,twocolumn,showkeys,showpacs,superscriptaddress]{revtex4}
\usepackage{epsfig,latexsym,amssymb}

\begin{document}

\title{Glassiness, Rigidity and Jamming of Frictionless Soft Core Disks}

\author{Daniel V{\aa}gberg}
\affiliation{Department of Physics, Ume{\aa} University, 901 87 Ume{\aa}, Sweden}
\author{Peter Olsson}
\affiliation{Department of Physics, Ume{\aa} University, 901 87 Ume{\aa}, Sweden}
\author{S. Teitel}
\affiliation{Department of Physics and Astronomy, University of
Rochester, Rochester, NY 14627}
\date{\today}

\begin{abstract}
The jamming of bi-disperse soft core disks is considered, using a variety of different protocols to produce the jammed state.  In agreement with other works, we find that cooling and compression can lead to a broad range of jamming packing fractions $\phi_J$, depending on cooling rate and initial configuration; the larger the degree of big particle clustering in the initial configuration, the larger will be the value of $\phi_J$.  In contrast, we find that shearing disrupts particle clustering, leading to a much narrower range of $\phi_J$ as the shear strain rate varies.  In the limit of vanishingly small shear strain rate, we find a unique non-trivial value for the jamming density that is independent of the initial system configuration.  We conclude that shear driven jamming is a unique and well defined critical point in the space of shear driven steady states.  We clarify the relation between glassy behavior, rigidity and jamming in such systems and relate our results to recent experiments.
\end{abstract}
\maketitle
 
 \section{Introduction}
 
An athermal system of hard or soft core interacting particles, for which thermal fluctuations are negligible (i.e. $T=0$), is found to undergo a {\it jamming transition} from a liquid-like state to a rigid but disordered solid as the packing fraction $\phi$ increases  above a critical value $\phi_J$ \cite{Jaeger}.   
Other physical systems similarly seem to show a transition from a liquid to a rigid but disordered state, as a function of some physical control parameter: foams change from flowing liquid to elastic solid once the applied shear stress $\sigma$ falls below a critical shear yield stress $\sigma_{\rm Y}$; supercooled liquids appear to freeze into a frozen glass as the temperature $T$ is lowered.  

\begin{figure}[h!]
\begin{center}
\includegraphics[width=2.0in]{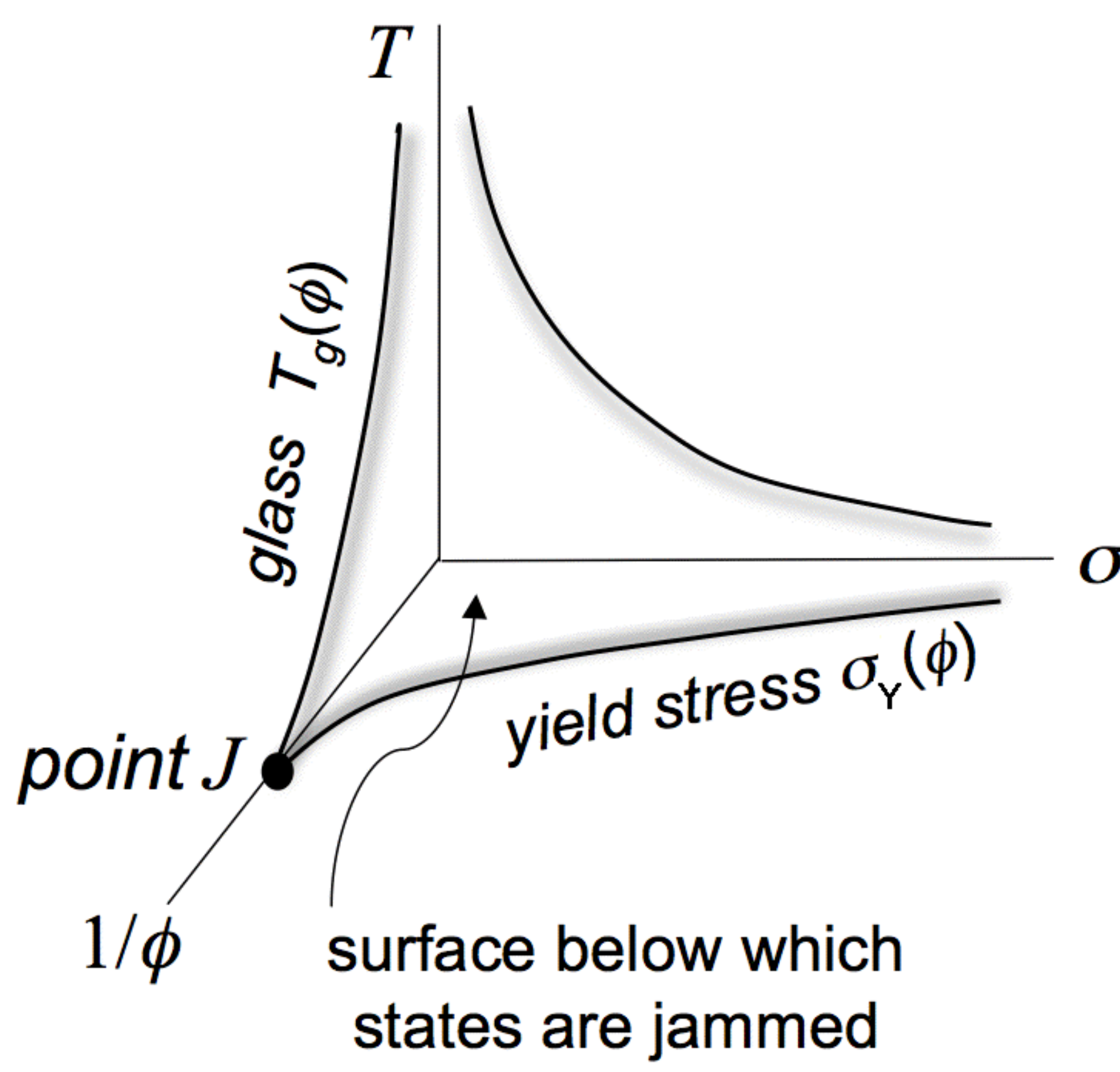}
\caption{Proposed jamming phase diagram, adapted from Ref.~\cite{OHern}.  Note that the axis along which packing fraction varies is $1/\phi$ rather than $\phi$.
}
\label{f0}
\end{center}
\end{figure}

These observations led Liu, Nagel and co-workers \cite{OHern,LiuNagel} to propose that these phenomena might be unified in terms of a three dimensional jamming phase diagram with the axes of packing fraction $\phi$, applied shear stress $\sigma$, and temperature $T$.  A surface in this three dimensional phase space separates jammed (i.e. rigid but disordered) from unjammed (i.e. liquid-like) states.  We sketch this jamming phase diagram in Fig.~\ref{f0}, following Ref.~\cite{OHern}.
As originally proposed \cite{OHern,LiuNagel}, this jamming surface represented points in phase space where the relaxation time $\tau$ of the system reached some experimentally defined large time scale.  The intersection of this surface with the equilibrium $(\phi,T)$ plane at $\sigma=0$ then describes the experimentally observed glass transition $T_{g}^{\rm exp}(\phi)$, where the viscosity of the liquid becomes immeasurably large upon cooling.  

For discussing possible critical behavior, it is of interest theoretically to consider the {\it critical} jamming surface that would correspond to a diverging time scale $\tau\to\infty$.  In this case the line $T_g(\phi)$ in Fig.~\ref{f0} would be a true {\it equilibrium} glass transition.  By saying {\it equilibrium} glass transition, we mean that $T_g(\phi)$ would locate a true singularity in the free energy, independent of the particular dynamics of the system.  Subsequently, when we refer to {\it equilibrium glass transition}, this will be what we mean.  Such a critical jamming surface would intersect the $\phi$ axis at $T=\sigma=0$ at a well defined point $\phi_0$, such that $T_g(\phi)\to 0$ as $\phi\to\phi_0$.  O'Hern et al. \cite{OHern} conjectured this point to be identical to the $T=0$ jamming transition of athermal particles, i.e. $\phi_0=\phi_J$, and denoted it ``point J".  Moreover, they conjectured that point J may act like a critical point and ``that it may control the region around it and thereby govern the nature of the entire jamming surface in the phase diagram"  \cite{OHern}.  Were this conjecture correct, then properties of the equilibrium glass transition would be intimately related to properties of athermal jamming.  In the following, we will denote ``point J" as the athermal jamming point, within some well defined physical protocol, where the packing fraction equals $\phi_J$ and $T=\sigma=0$.

It must be noted that in models of simple liquids, such as those considered in this work, the existence of an equilibrium glass transition at a {\it finite} temperature, $T_g(\phi)>0$, remains a much debated question.  Works have suggested that the critical jamming surface, where $\tau\to\infty$, may collapse entirely into the $T\to 0$ plane. 
See Ref.~\cite{Coniglio} for a recent review.  However, even in this case, one might expect at $\sigma=0$ a line of zero temperature glass transitions, that would terminate at some specific lowest packing fraction $\phi_0$, with viscosity diverging as $T\to 0$ for all $\phi\ge\phi_0$.  If such a $\phi_0$ was identical to the $\phi_J$ of point J, one would again have a connection between equilibrium glassy behavior (albeit a $T_g\to 0$ glass transition) and athermal jamming.  

Although the above conjecture is appealing, several recent works have suggested that the actual situation may be more complicated.  
Some simulations have suggested that an equilibrium glass transition may be controlled by a different critical point, sometimes referred to as ``point G'', that is distinct from the athermal jamming point J. 
Equilibrium simulations of {\it hard} spheres (where temperature $T$ plays no role and density $\phi$ is the only parameter) in three dimensions (3D) by Brambilla et al. \cite{Brambilla} claim a glass transition at a $\phi_0$ that is distinctly lower than the typical values of $\phi_J$ obtained from compression; the reduced pressure $p/T$ remains finite at $\phi_0$, in contrast to the athermal jamming of hard spheres where $p/T$ is expected to diverge.  Equilibrium simulations of soft spheres in 3D by Berthier and Witten \cite{Berthier1,Berthier2} show a scaling in the $(\phi,T)$ plane that similarly suggests a $T\to 0$ glass transition at a $\phi_0$ lower than the athermal jamming $\phi_J$.  Similar results have been suggested by Xu et al. \cite{Xu}.  Other works by Donev et al. \cite{Donev} argue that in a properly equilibrated hard sphere system there is no glass transition, unless some constraint is imposed to prevent crystallization.

At the same time, recent works have illustrated that the precise value $\phi_J$ of the athermal point J is not unique, even within a given specified model, but can depend upon the particular protocol used to prepare the jammed state. Donev et al. \cite{Donev} show, for both frictionless mono-disperse spheres in 3D and bi-disperse disks in two dimensions (2D), that  for compression driven jamming, $\phi_J$ depends on the rate of compression.  
Chaudhuri et al. \cite{Berthier3} show that when compressing configurations equilibrated at an initial $\phi_{\rm init}$,  bi-disperse frictionless spheres in 3D jam at a $\phi_J$ that increases with $\phi_{\rm init}$; hence the value of $\phi_J$ can depend not only on the compression rate, but on the particular initial configuration  from which one starts the compression.   
Recent theoretical works \cite{Krzakala,Parisi} on mean-field hard sphere models have found similar results: a continuous range of athermal jamming densities at infinite $p/T$, with a $\phi_J$ that varies according to compression or cooling rate, as well as a distinct equilibrium glass transition at a finite $p/T$.

In this work we consider a variety of jamming protocols for a 2D system of frictionless bi-disperse soft-core disks, focusing on protocols which do not involve compression.  We also carry out equilibrium Monte Carlo (MC) simulations of hard disks to look for the onset of glassy behavior prior to jamming, i.e. at a $\phi_0<\phi_J$, as was previously observed in 3D.  Our main conclusion is that, unlike jamming driven by compression or cooling, athermal shear driven jamming results, in the limit of a vanishingly small shear strain rate $\dot\gamma\to 0$, in a unique, well-defined, non-trivial, value of $\phi_J$ that is independent of the system's starting configuration.  This result follows from our observation that shearing breaks up the clustering of big particles that can lead to phase separation and crystallization under slow compression or cooling.
The remainder of this paper is organized as follows.
In Section II we describe our model of bi-disperse frictionless soft core disks, give our precise procedure for determining the jamming packing fraction $\phi_J$, and describe the different jamming protocols we will consider.  In Section III we present our numerical results.  In Section IV we relate our results to some recent experiments and summarize our conclusions.

\section{Model}

Our system is a bi-disperse mixture of frictionless disks with diameter ratio $d_{\rm B}/d_{\rm S}=1.4$, as has been used in earlier works \cite{OHern,Olsson}.  The fraction of bigger particles is $x_{\rm B}=1/2$.  
The disks interact via a pairwise soft-core repulsive contact interaction, that is harmonic in the particle overlap,
\begin{equation}
V(r_{ij})=\left\{ 
\begin{array}{ll}\epsilon(1-r_{ij}/d_{ij})^2/2 &{\rm for}\quad r_{ij}<d_{ij}\\ 0 &{\rm for}\quad r_{ij}\ge d_{ij}\end{array}\right.
\end{equation}
where $r_{ij}$ is the distance between the centers of two particles $i$ and $j$, and $d_{ij}$ is the sum of their radii. Length is measured in units such that the smaller diameter is unity, and energy is measured in units such that $\epsilon=1$.  A system of $N$ disks in an area $A$ thus has a packing fraction (density),
\begin{equation}
\phi = N\pi(0.5^2+0.7^2)/(2A)\enspace.  
\label{erho}
\end{equation}
We use a simulation cell of area $A=L_xL_y$, with equal length and width, $L_x=L_y$.

Starting from a set of physically motivated initial states at fixed  $\phi$, we use a non-linear conjugate gradient method to quench each state to its local energy minimum (the {\it inherent structures}).  
For soft core particles, mechanically stable states exist at values of $\phi$ above the jamming $\phi_J$.  Such states are characterized by a finite interaction energy, pressure, and shear yield stress; all these vanish as $\phi\to\phi_J$ from above \cite{OHern}.  
Energy minimized states with an energy per particle below a certain fixed very small threshold value, $E/N<e_{\rm cut}$, are therefore regarded as unjammed; otherwise the state is considered to be jammed.  This criterion for jamming \cite{OHern} has been shown \cite{Donev1} to be essentially the same as ``strictly jammed" in the classification scheme of Torquato and Stillinger \cite{TorqStill}.  
In this manner we count the fraction $f(\phi)$ of these energy minimized states which are {\it jammed}.

As $\phi$ increases, $f(\phi)$ varies rapidly from zero to unity, with $f(\phi)$ approaching a sharp step function as the number of particles $N\to\infty$.  The location of the step then determines the jamming packing fraction  $\phi_J$ {\it for that initial set of states}.  We consider two classes of initial states:  (i) Equilibration at a finite $T$, which may be thought of as the equilibrium temperature of a glassy system, or as an effective temperature of kinetic motion in a granular system with uniform mechanical agitation.  Quenching corresponds to suddenly turning off the agitation and allowing the system to relax.  In the limit $T=\infty$ one chooses random initial positions.   This is the ensemble studied by O'Hern et al. \cite{OHern} and we will denote it as ``RAND."  (ii) Shearing at a constant uniform shear strain rate $\dot\gamma$.  Quenching corresponds to suddenly turning off the shear and allowing the system to relax to a mechanically stable or to an unjammed state.  The limit $\dot\gamma\to 0$ gives {\it quasistatic} shearing (``QS"), as studied previously by Heussinger and co-workers \cite{Heussinger, Heussinger2}.  

The specific conjugate gradient algorithm we use to energy minimize is the Polak-Ribiere method \cite{NR}.
We stop the energy minimization when one of the following conditions is met: (i) the relative decrease in the energy $\Delta E/E$ after 50 iterations is smaller than $10^{-10}$, or (ii) the energy per particle falls below a certain small threshold value, $E/N < e_{\rm cut}$. In the second case, we consider the state to be {\it unjammed}.  We find that the threshold value $e_{\rm cut}=10^{-16}$ gives a clear separation between the jammed and unjammed states up to the largest system size we have studied (see appendix A for further details).

\section{Results}

\subsection{Random vs Quasistatic Shearing Ensembles}

\begin{figure}[h!]
\begin{center}
\includegraphics[width=3.3in]{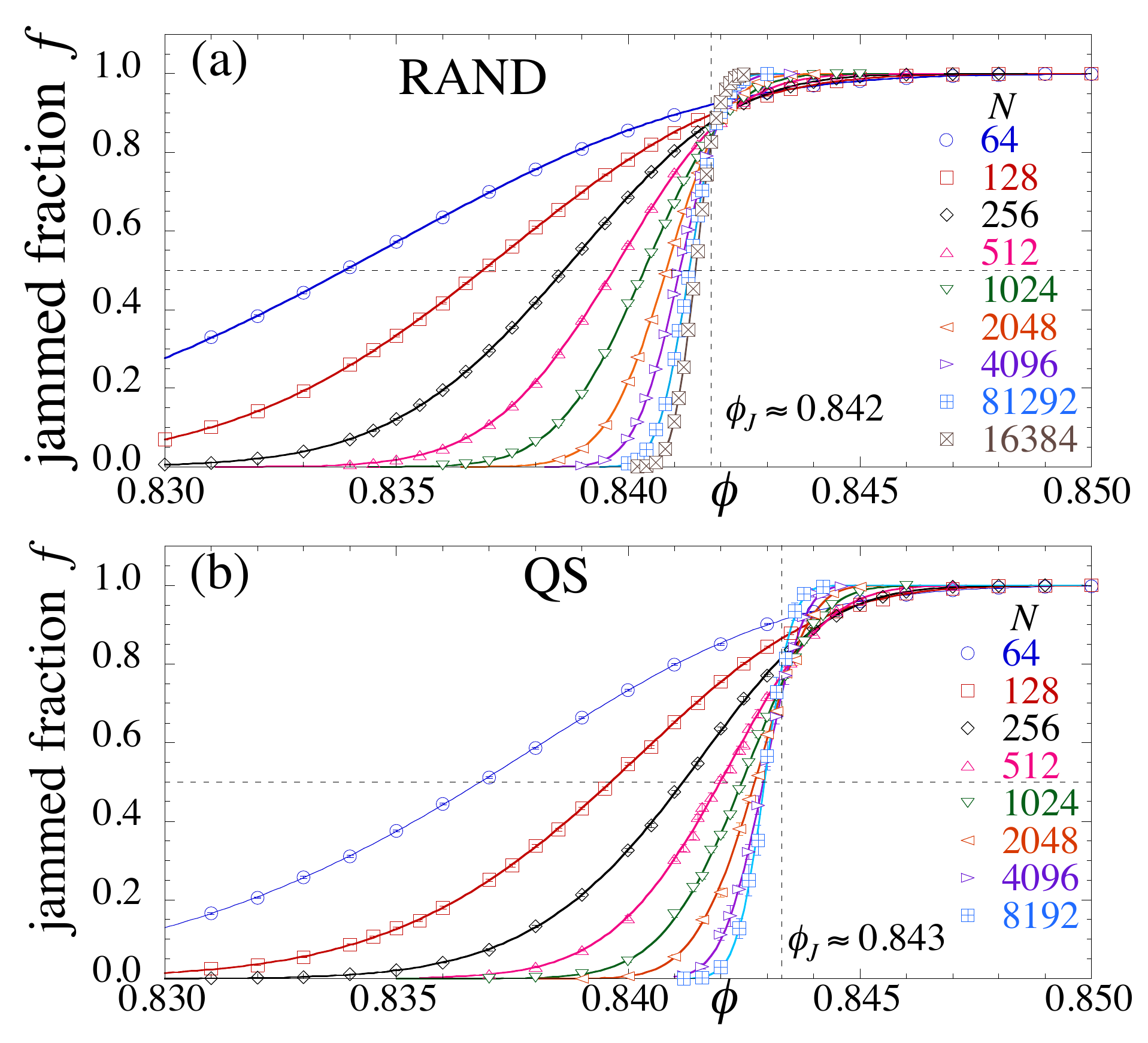}
\caption{(color online) Jammed fraction $f$ vs packing fraction $\phi$ for systems of different numbers of particles $N$.  (a) and (b) are for the RAND and QS ensembles, respectively.  Vertical dashed lines indicate the limiting $N\to\infty$ value of the jamming density $\phi_J$ in each case.
}
\label{f1}
\end{center}
\end{figure}

In Fig.~\ref{f1} we plot our results for the RAND and QS ensembles, showing how the jammed fraction $f(\phi)$ sharpens to a step function as the number of particles $N$ increases.  For RAND we average over at least 10000 initial configurations for each value of $\phi$.  
For QS we average over $10-20$ independent runs, each sheared a total strain $\gamma \sim 4-8$ for our biggest size, but $\gamma=200$ for our smallest size.  
We use Lees-Edwards boundary conditions \cite{LeesEdwards} to model uniform shear strain, energy minimizing after each small strain step $\Delta\gamma$. For $N\ge4096$ we use $\Delta\gamma=10^{-5}$, while for $N<4096$ we use $\Delta\gamma=10^{-4}$.  We have explicitly verified that for all, but perhaps the biggest size $N=8192$, our value of $\Delta\gamma$ is small enough not to influence our results (see appendix A for details).  We clearly see that the two ensembles approach different jamming densities, $\phi_J^{\rm RAND}\simeq 0.842$ while $\phi_J^{\rm QS}\simeq 0.843$ \cite{note2}.  
 
\subsection{Equilibration at Finite Temperature}

\begin{figure}[h!]
\begin{center}
\includegraphics[width=3.3in]{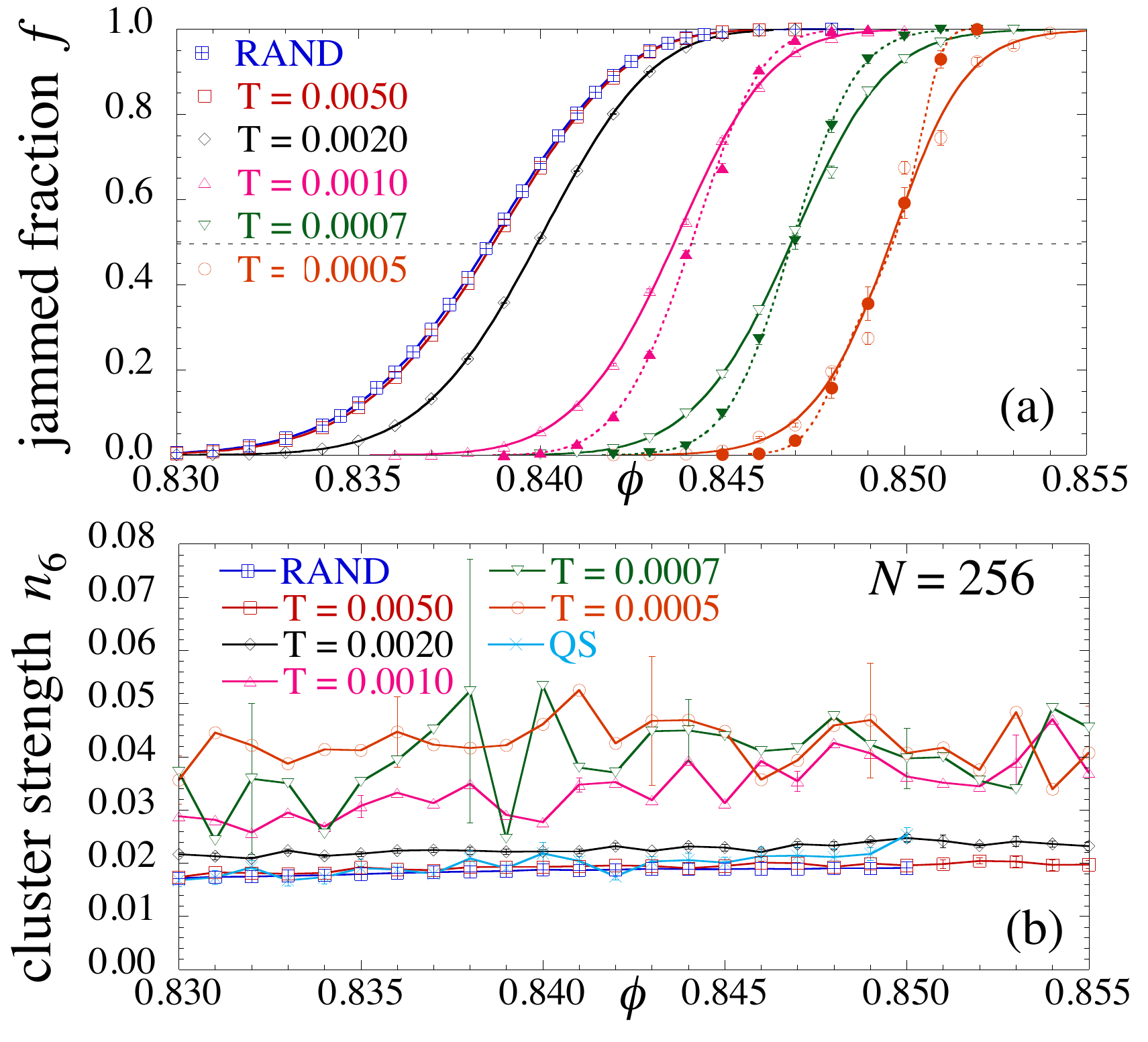}
\caption{(color online) (a) Jammed fraction $f$ vs packing fraction $\phi$ comparing RAND with ensembles quenched after thermal equilibration at a fixed temperature $T$.  Open symbols (solid lines) are for $N=256$ particles, while closed symbols (dashed lines) are for $N=512$ particles. (b) Clustering strength parameter $n_6$ vs $\phi$ for the cases shown in (a) as well as for QS.  Representative error bars are shown at select data points at the lowest $T$'s. }
\label{f2}
\end{center}
\end{figure}

Next we consider initial states equilibrated at a fixed temperature $T$.  In Fig.~\ref{f2}a we plot the jammed fraction $f(\phi)$ resulting from the energy minimized states arising from these thermally equilibrated initial states, comparing 
RAND with several finite values of $T$, for $N=256$ particles.  For the three lowest $T$ we also show results for $N=512$ to illustrate that increasing $N$ continues to lead to a sharpening of the transition as seen in Fig.~\ref{f1}.  To equilibrate at $T$ we do ordinary MC simulations, at each step displacing a randomly chosen particle by a random amount and accepting or rejecting the move according to the Metropolis algorithm.  $N$ attempted particle moves is 1 MC pass. 
At our lowest $T$, we use 10 independent runs, each with roughly $10^{8}$ MC passes.
We judge that we have equilibrated when particles have, on average, diffused a distance equal to a few particle diameters.  We see that $T=5\times 10^{-3}$ is essentially equal to the $T=\infty$ ensemble RAND, but that as $T$ decreases, $\phi_J(T)$
increases.  Our lowest  $T^*=5\times 10^{-4}$ gives our largest $\phi_J(T^*)\simeq 0.850$ \cite{note1}. For such high densities, our runs at $T^*$  are just at the border of equilibrating; we would need much longer runs to try and equilibrate at even lower $T$. Similar results have recently been found for continuous cooling with different fixed rates \cite{OHern2}.   These results are in good agreement with recent predictions from a mean-field-like hard-sphere model \cite{Krzakala}.

Roughly the same range of $\phi_J$ was found by Donev et al. \cite{Donev} from slow compression of bi-disperse hard disks (they use $x_{\rm B}=1/3$).  
Donev et al. argue that an increased $\phi_J$ results from an increased order due to the clustering of big particles.  To check for such clustering we have computed the fraction, $n_6$, of big particles which have 6 nearest neighbors (as determined by Delaunay triangularization) that are also big particles.  In Fig.~\ref{f2}b  we plot $n_6$ vs $\phi$ for the cases of Fig.~{\ref{f2}a, as well as QS.  We see little difference in $n_6$ comparing RAND, QS, and the highest $T$, however $n_6$ systematically increases as $T$ decreases.  The increasing fluctuations in $n_6$ as $\phi$ varies at low $T$ reflect the increasing difficulty to equilibrate.
Donev et al. have argued that, given sufficiently long equilibration, even higher values of $\phi_J$ might be achieved, up to the maximum  $\phi_{\rm max}\simeq 0.91$ of fully phase separated  lattices of big and small particles.  We expect a similar situation in our system, {\it if} we could equilibrate at even lower $T$.   

\subsection{Shear Driven Steady States}

Next we consider shearing the system.  For our simulations at a finite shear strain rate $\dot\gamma$ we use Durian's \cite{Durian} foam dynamics: overdamped motion with viscous damping to the local average shear flow velocity.  For shear flow in the $\hat x$ direction we have,
\begin{equation}
 \frac{d{\bf r}_i}{dt} = -{C}\sum_j\frac{dV({\bf r}_{ij})}{d{\bf r}_i} + y_i \dot\gamma\; \hat{x}\enspace,
 \label{em}
\end{equation} 
where the last term $y_i\dot\gamma\;\hat x$ is the average shear flow veloctiy.  Lees-Edwards boundary conditions are used, and we choose units of time such that $C=1$.
We run the simulations up to a certain total strain $\gamma=\dot\gamma t$ ($\gamma=33$ for our smallest $\dot\gamma$), and then sampling configurations uniformly from the resulting shear flow, we energy minimize them and count the resulting fraction that are jammed.  
In Fig.~\ref{f3} we show results for the jammed fraction $f(\phi)$ for $N=256$ particles.  

We stress that the energy minimization step, representing a sudden switch off of the applied strain rate $\dot\gamma$, is crucial to this calculation.  Were we to sample the steady state distribution  at fixed $\dot\gamma$ directly, {\it all} states would have a finite energy yet all states are flowing; our criterion for  computing the jammed fraction is only applicable to mechanically stable states at rest.  Thus the $\phi_J(\dot\gamma)$ determined by the present procedure should {\it not} be viewed as representing a jamming transition for driven steady states at finite $\dot\gamma$ (there is none, since all such states by definition are flowing); rather it represents the jamming point resulting from a particular dynamic protocol for creating statically jammed states, i.e. relaxation to rest from initial states driven at a finite shear rate $\dot\gamma$.

Our fastest shear rate $\dot\gamma=10^{-3}$ gives results equal to the random initial positions of RAND.  From this we can infer that, in a rapidly sheared system, the soft core interactions are playing little role in ordering the particles, and the system is thus passing through effectively random configurations.
Our slowest shear rate $\dot\gamma=10^{-8}$ is clearly converging to the QS limit.  Thus quasistatic shearing, in which the system is always instantaneously relaxing into its nearest local energy minimum as it is slowly sheared, is the appropriate $\dot\gamma\to 0$ limit of the overdamped dynamics of Eq.~(\ref{em}).  From our results in Figs.~\ref{f1}b and \ref{f3} we thus conclude that there is indeed a well defined jamming density $\phi_J^{\rm QS}$ in the $\dot\gamma\to 0$, $N\to\infty$, limit.


\begin{figure}[h!]
\begin{center}
\includegraphics[width=3.4in]{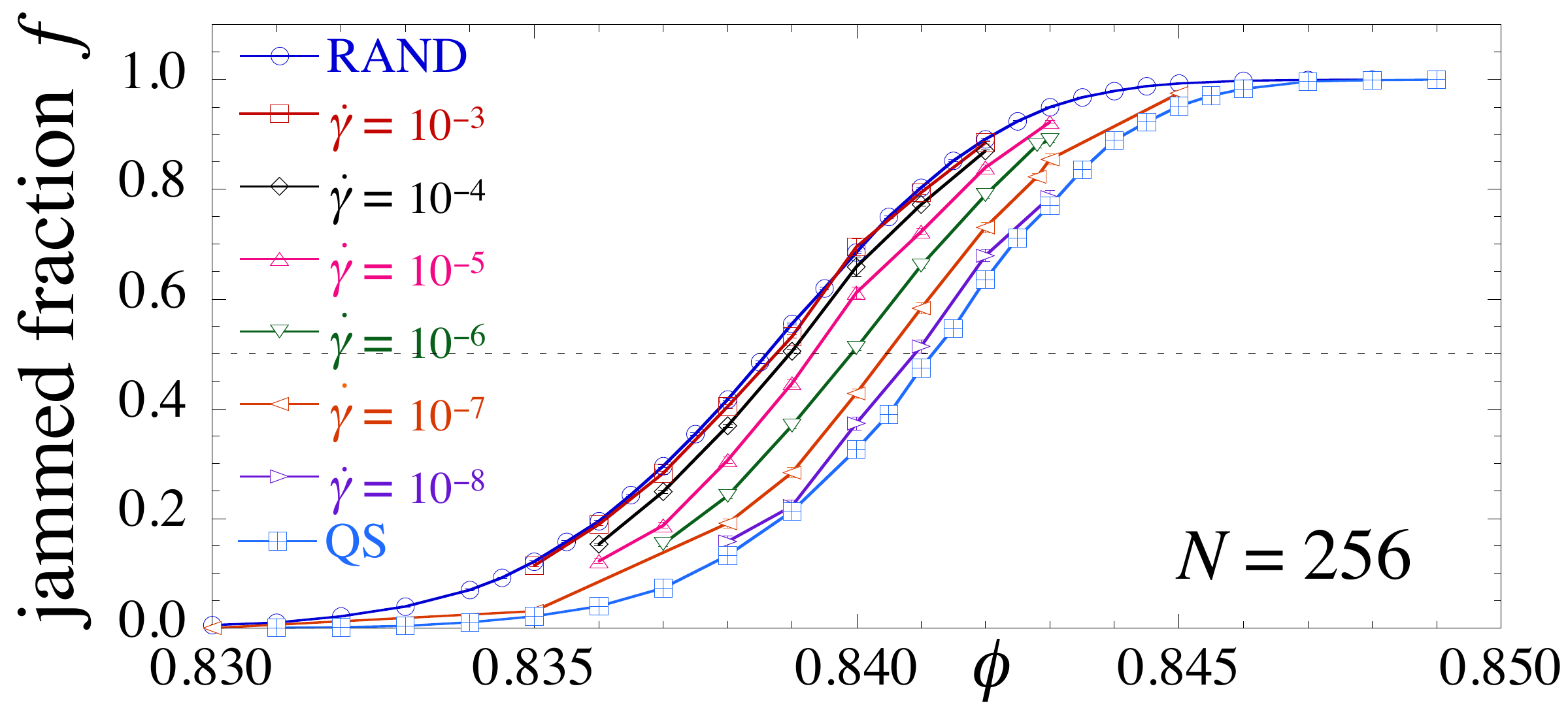}
\caption{(color online) Jammed fraction $f$ vs packing fraction $\phi$ for a system of $N=256$ particles, comparing RAND and QS ensembles with ensembles quenched from a fixed finite shear strain rate $\dot\gamma$.  }
\label{f3}
\end{center}
\end{figure}

\subsection{Hard Disk Equilibrium Monte Carlo}

Next we consider the dynamic behavior at low densities $\phi<\phi_J^{\rm RAND}$.  Following Brambilla et al. \cite{Brambilla}, we simulate the diffusion of $N=1024$ {\it hard} core disks using local MC moves in which a randomly selected particle is displaced a random amount within a box of length $0.1d_{\rm S}$ about its center; a move is accepted only if the non-overlap hard disk constraint is obeyed.  $N$ such attempted moves  corresponds to one MC pass, which we equate to one unit of time.
Measuring the self-part of the intermediate scattering function \cite{Brambilla}, 
\begin{equation}
F_s(q,t)\equiv{1\over N}\langle\sum_i {\rm e}^{i{\bf q}\cdot[{\bf r}_i(t)-{\bf r}_i(0)]}\rangle\enspace,
\end{equation}
with ${\bf q}=(2\pi/1.2d_{\rm S})\hat x$, we define the relaxation time $\tau$ by $F_s(q,\tau)=1/e$.  In Fig.~\ref{f4}a we show results for $\tau$ vs $\phi$.  At low $\phi\lesssim 0.76$ equilibration is relatively straight forward.  At larger $\phi$, we use the following procedure to try and stay on the metastable glassy part of the equation of state: starting from the ending configuration of the previous value of $\phi$,  we compress the system an amount $\Delta\phi=0.005$, and then simulate for a time of roughly $100\tau$ before increasing $\phi$ again.  At our lowest $\phi$ this corresponds to $3\times 10^{5}$ MC passes; for our highest $\phi$ this is $5\times 10^8$ MC passes.
We leave aside the question whether $\tau(\phi)$ is truly diverging at an {\it ideal glass transition} $\phi_0$, as suggested by Berthier and Witten \cite{Berthier1,Berthier2}, or whether the growth in $\tau$ is a kinetic effect of falling out of equilibrium, as argued by Donev et al. \cite{Donev}.
Here we just note that $\tau$ clearly grows many orders of magnitude by the time one reaches $\phi_0\sim 0.80<\phi_J^{\rm RAND}\simeq 0.842$, thus leading to glassy behavior before the onset of our lowest jamming density.  In Fig.~\ref{f4}b we show the corresponding cluster strength  $n_6$, which we see increases rapidly with increasing $\phi$.  

In their work, Chaudhuri et al. \cite{Berthier3} observed that when they equilibrated their system first as hard spheres at some initial $\phi_{\rm init}$, and then slowly compressed them to reach jamming, the $\phi_J$ that resulted increased with increasing $\phi_{\rm init}$.  From Fig.~\ref{f4}b we see that, for equilibrated systems, the clustering strength $n_6$ increases rapidly with increasing $\phi$.   The results of Chaudhuri et al. are thus  consistent \cite{note} with the assertion by Donev et al. \cite{Donev} that initial configurations with greater clustering result in jamming at larger $\phi_J$, when compressed.

\begin{figure}[h!]
\begin{center}
\includegraphics[width=3.4in]{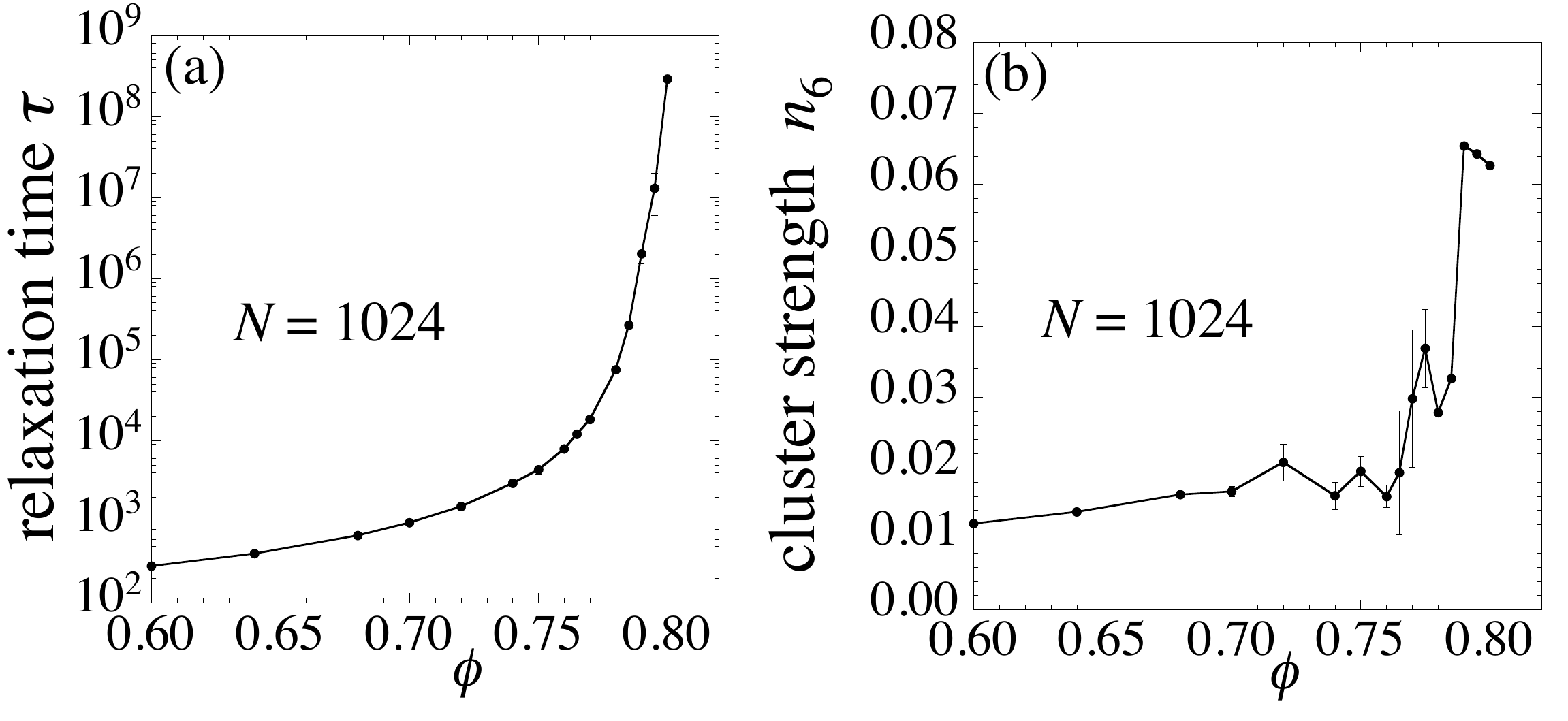}
\caption{(a) Relaxation time $\tau$ and (b) clustering strength parameter $n_6$ vs $\phi$ for $N=1024$ diffusing hard core particles.  Representative error bars are shown.}
\label{f4}
\end{center}
\end{figure}

The data of Fig.~\ref{f4} does not represent true equilibrium at the largest values of $\phi$ shown.  We have found that we are able to more properly equilibrate the system if we include non-local swaps between big and small particles in our MC moves.  Such moves are, of course, unphysical when modeling a continuous dynamics of the particles, but they are perfectly acceptable for sampling true equilibrium.  With such moves we find evidence for a transition near $\phi\sim 0.78$ to a phase separated coexistence between a liquid mixture of big and small particles and a solid of big particles, just as was predicted by Donev et al. \cite{Donev}.  In such true equilibrium states, $n_6$ becomes even larger than found in Fig.~\ref{f4}b.  

\subsection{Compression vs Shear Driven Jamming}

To illustrate our above results on jamming, we next consider the following numerical ``experiment".
Since the largest values of $n_6$ in Fig.~\ref{f4}b are  slightly larger than found in Fig.~\ref{f2}b from cooling soft disks, we expect that compression of configurations equilibrated at densities $\phi\sim 0.80$ should result in relatively high jamming densities.  We therefore take one configuration at $\phi=0.80$, sampled from the states that produced the data of Fig.~\ref{f4}; we denote this as configuration ``A".  We take a second configuration ``B", obtained also at $\phi=0.80$, but after doing MC with particle swaps so as to achieve a better equilibration of the system and a higher degree of particle clustering.  Configuration A has $n_6=0.037$ while B has $n_6=0.168$.  Both have $N=1024$ particles.  We show these initial configurations A and B in Fig.~\ref{f5}.

\begin{figure}[h!]
\begin{center}
\includegraphics[width=3.3in]{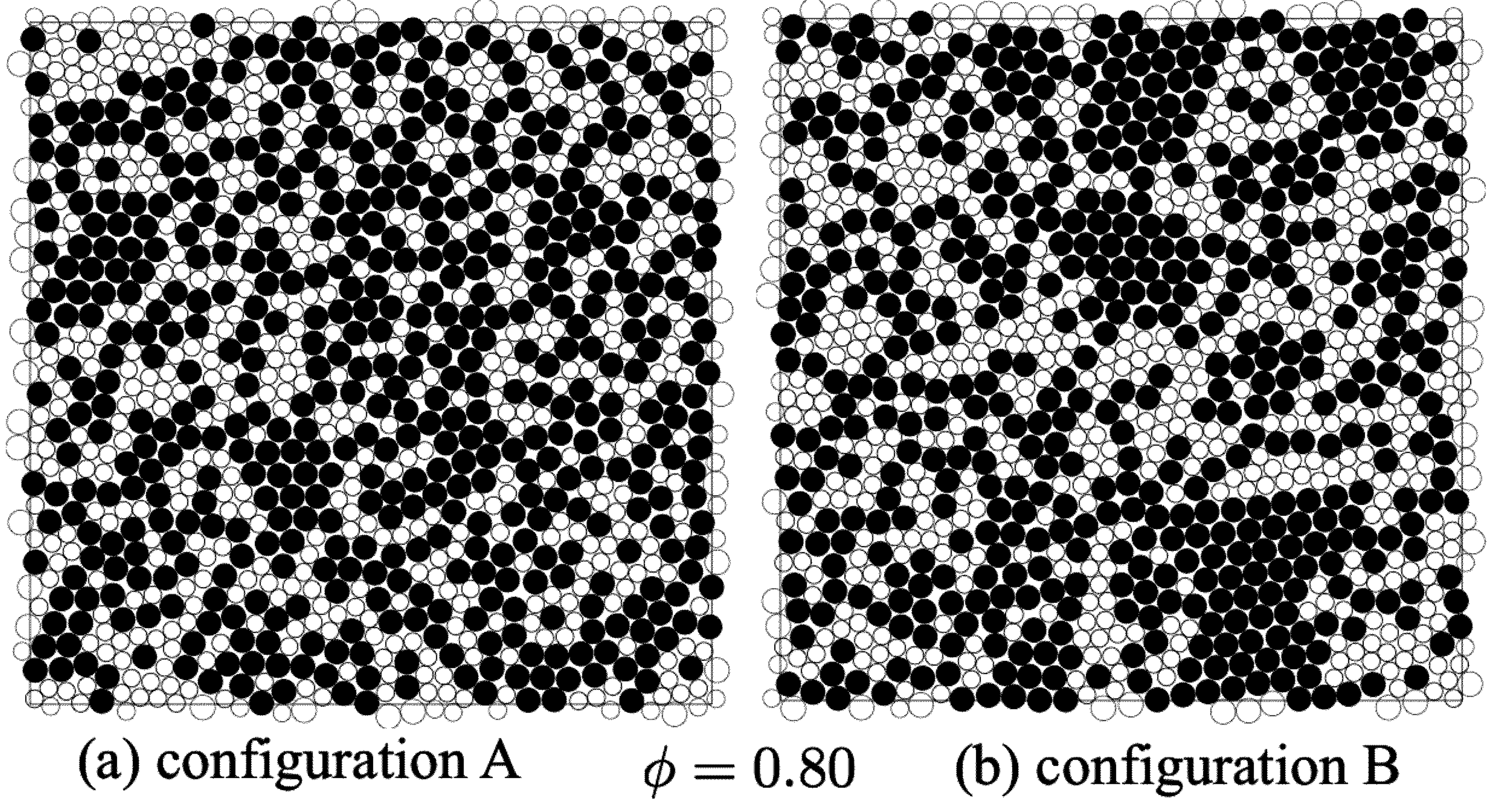}
\caption{Initial unjammed configurations A and B at $\phi=0.80$. Configuration B is seen to have a greater degree of big (black) particle clustering than A. $N=1024$.}
\label{f5}
\end{center}
\end{figure}

We then uniformly compress both configurations A and B in steps of $\Delta\phi=10^{-4}$, relaxing the system to its local energy minimum after each compression step.  In Fig.~\ref{f6}a we plot the resulting energy per particle $E/N$ vs $\phi$.  We see that A jams at the relatively high value of $\phi_J^{\rm A}\simeq 0.8534$, while B jams at the even higher $\phi_J^{\rm B}\simeq 0.8559$.  We then return to these configurations as they were  at $\phi=0.85$.  Because $0.85$ is below the jamming density of either system, these are unjammed, stress-free, states.  We now quasistatically shear these configurations using a strain step $\Delta\gamma=10^{-4}$.  Our results are shown in Fig.~\ref{f6}b.  We see that after relatively small strains of $\gamma=0.05$ for A, and $\gamma=0.077$ for B, both systems jam.  This is as expected since $\phi=0.85>\phi_J^{\rm QS}$.  In Fig.~\ref{f6}c we plot the instantaneous value of $n_6$ for these two configurations, as a function of total shear strain $\gamma$ at $\phi=0.85$, showing results out to a much larger total strain $\gamma=35$ than in Fig.~\ref{f6}b.  We see that after a certain amount of shearing, $n_6$ for both A and B drop down to the values typical of the QS ensemble (see Fig.~\ref{f2}b).  We thus conclude that shearing breaks up the particle clustering that can lead to high jamming densities under slow compression, and that this effect is responsible for the well defined jamming density $\phi_J^{\rm QS}$ in the limit of vanishingly small shear rate, $\dot\gamma\to 0$.

\begin{figure}[h!]
\begin{center}
\includegraphics[width=3.3in]{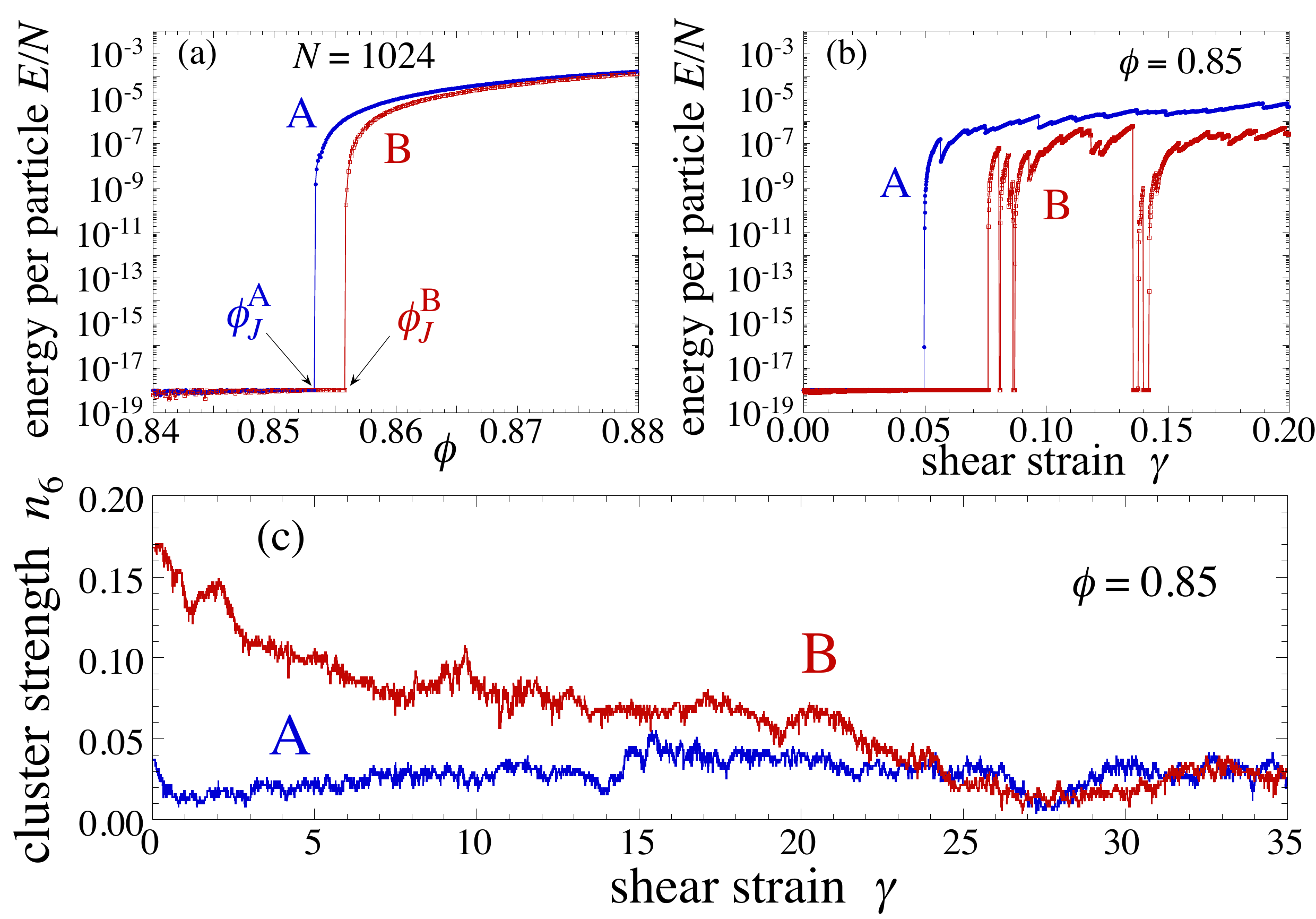}
\caption{(color online) (a) Energy $E/N$ vs $\phi$ for the two configurations A and B of Fig.~\ref{f5} undergoing uniform compression.  (b) $E/N$ vs strain $\gamma$ for the same configurations undergoing uniform shear starting from stress-free states at $\phi=0.85$. (c) Clustering strength $n_6$ vs $\gamma$ as the two configurations are sheared at $\phi=0.85$. $N=1024$.}
\label{f6}
\end{center}
\end{figure}

To further illustrate this point, in Fig.~\ref{f7} we plot the {\it strain averaged} cluster strength,
\begin{equation}
\langle n_6\rangle_\gamma\equiv {1\over\gamma}\int_0^\gamma d\gamma^\prime n_6(\gamma^\prime)\enspace,
\end{equation}
the strain averaged energy per particle $\langle E/N\rangle_\gamma$, and the strain averaged jammed fraction $\langle f\rangle_\gamma$ vs total shear strain $\gamma$, as we quasistatically shear three different initial states: the moderately clustered state A and the highly clustered state B, as in Figs.~\ref{f5} and \ref{f6}, and a state C that starts with particles in random positions.  

In Fig.~\ref{f7}a,b,c, we show results at the density $\phi=0.85> \phi_J^{\rm QS}$ (the same density as in Fig.~\ref{f6}c).  
In Figs.~\ref{f7}d,e,f we show results at the lower density $\phi=0.84<\phi_J^{\rm QS}$.
In both cases we see that as $\gamma$ increases, the strain averaged quantities for the different initial configurations approach a common steady value.  For the moderately clustered initial state A, and the random initial state C, this happens after a relatively short strain; for the highly clustered initial state B, it takes considerably longer to lose memory of the initial state.  Note, for $\phi=0.84$, $f\approx 0.05$ and so the system is rarely jammed; as a finite energy comes only from jammed configurations, the rarity of jammed configurations at the low $\phi=0.84$ is the reason for the larger fluctuations observed in the curves for $\langle E/N\rangle_\gamma$ shown in Fig.~\ref{f7}e.

These observations illustrate two important points: (i) quasistatic shearing over long total strains produces a well defined ensemble of states that is independent of the initial configuration, and (ii) the process of shearing, no matter how slow, destroys the clustering that can produce large $\phi_J$'s under compression.  
It is for this reason that $\phi_J^{\rm QS}$ represents a true, well-defined, jamming transition in the limit of vanishingly small shear strain rate $\dot\gamma\to 0$, and does not suffer from the questions of equilibration and protocol that jamming from compression or cooling does.

\begin{widetext}

\begin{figure}[h!]
\begin{center}
\includegraphics[width=7in]{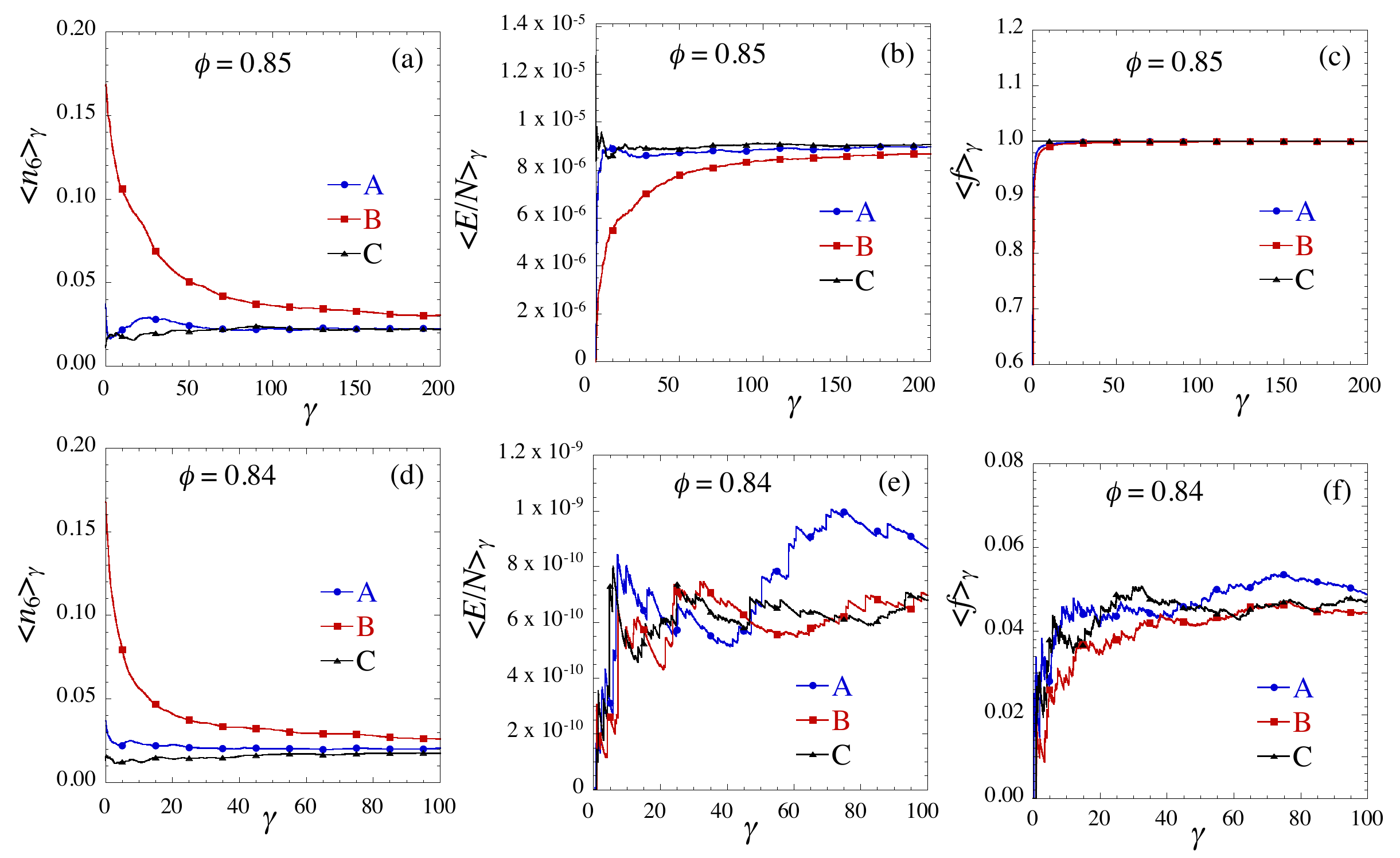}
\caption{(color online) (a) Strain averaged cluster strength $\langle n_6\rangle_\gamma$, (b) energy per particle $\langle E/N\rangle_\gamma$, and (c) jammed fraction $\langle f\rangle_\gamma$, vs total strain $\gamma$ at $\phi=0.85>\phi_J^{\rm QS}$;  plots (d), (e), and (f) are the same quantities at the lower density $\phi=0.84<\phi_J^{\rm QS}$.
The three curves correspond to simulations starting from three different initial configurations:  A is the moderately clustered configuration, and B is the highly clustered configuration, of Figs.~\ref{f5} and \ref{f6}, while C is a configuration with random initial particle positions. The system has $N=1024$ particles.  Note, symbols are displayed on every $2000^{th}$ data point for $\phi=0.85$, and every $1000^{th}$ data point for $\phi=0.84$, to help with curve identification.}
\label{f7}
\end{center}
\end{figure}

\end{widetext}

\section{Discussion and Conclusions}

We can relate our results to two recent experiments. Lechenault et al. \cite{Dauchot}, in experiments on vibrated bi-disperse brass disks, interpret their results in terms of three relevant densities, $\phi_g<\phi_J<\phi_a$ (see their Fig.~1). $\phi_g$ they call the  {\it glass/jamming transition} where the structural relaxation time rapidly grows large on experimental time scales; we can identify this with the behavior in our Fig.~\ref{f4}a.  At $\phi_a$ they say that the system reaches the {\it fully arrested
state}; we can identify this as the relatively large jamming density one can obtain from slow compression.  For $\phi_g<\phi<\phi_a$, they say ``strong vibration can still induce micro-rearrangements through collective contact slips that lead to a partial exploration of the portion of phase space, restricted to a particular frozen structure" and they find a diverging time and length scale at a $\phi_J\simeq 0.842$ within this region; they refer to this $\phi_J$ as the {\it rigidity/jamming transition}.  We believe this is the region where small shear displacements remain possible (as illustrated in Fig.~\ref{f6}b at low $\gamma<0.05$) and that their $\phi_J$ corresponds to the $\phi_J^{\rm QS}$ of quasistatic shearing.
In another work by Zhang et al. \cite{Behringer}, a system of disks was prepared in a stress-free configuration at a density $\phi=0.758$, but upon shearing at constant $\phi$, the system jammed relatively quickly.  The comparatively low value of $\phi$ in these experiments, as well as the low average contact number $Z\sim 3$ they find at jamming, suggests, as the authors say, that friction is playing an important role in these experiments.  Here we point out, however, that exactly the same behavior may be observed in frictionless disks, as illustrated by our Fig.~\ref{f6}.

To conclude, we have considered various approaches to the jamming of 2D disks.  Consistent with earlier works, our results in Figs.~\ref{f2}a and \ref{f6}a show that a relatively wide range of jamming densities $\phi_J$ are possible when compressing mechanically stable configurations or when cooling thermally equilibrated configurations.  We can view compression and cooling as {\it quasi-equilibrium} processes, since they involve changes in the equilibrium variables of $\phi$ and $T$.  We see that, rather than jamming being defined at a unique density, the value of $\phi_J$ from such processes is affected by details of the specific protocol, such as compression or cooling rate, the relative degree of order (particle clustering) in the initial configurations one starts from, and presumably other details of the compression or cooling algorithm.  For infinitesimally slow compression or cooling, it remains unclear if there is a well-defined limiting value of $\phi_J$ that is lower than the $\phi_{\rm max}\simeq 0.91$ of fully phase separated  close packed lattices of big and small particles.
These observations suggest that there is no disordered athermal jamming point that is the well defined $T\to 0$ limit of an {\it equilibrium} glass transition $T_g(\phi)$ in the $(\phi, T)$ plane.  Such an equilibrium glass transition, if it exists, must by definition be protocol independent, whereas athermal jamming via compression or cooling appears not to be.  

However, if one follows Liu and Nagel \cite{LiuNagel}, and moves off equilibrium into the phase space of shear driven {\it nonequilibrium} steady states, then we find that there is a {\it unique} well-defined athermal jamming transition in the phase space $(\phi,T,\dot\gamma)$, located at $(\phi_J^{\rm QS}, T=0,\dot\gamma\to 0)$ (note, it is more convenient here to think of the nonequilibrium axis as being the shear strain rate $\dot\gamma$ rather than the shear stress $\sigma$).  We stress that the $(\phi, T)$ plane at $\dot\gamma\to 0$ is {\it not} the plane of equilibrium; rather it is the plane of quasistatically sheared steady states.  This is most easily seen by noting that for $\phi>\phi_J^{\rm QS}$, the quasitatically sheared system has a finite average shear stress $\sigma$ (which is just the dynamic yield stress $\sigma_{\rm Y}(\phi)$), whereas in equilibrium one would expect to have $\sigma=0$.  Unlike with compression or cooling, the limit of infinitesimally slow shearing gives a well defined value $\phi_J^{\rm QS}$ clearly less than $\phi_{\rm max}$.  Unlike compression or cooling, the value of $\phi_J^{\rm QS}$ is independent of the initial configuration one starts from; the process of quasistatic shearing creates a well defined ensemble that is independent of the starting configuration.  This is illustrated by the results of our Figs.~\ref{f6}--\ref{f7} where, even starting from a carefully prepared dense unjammed state with a large degree of particle clustering, we find that quasistatic shearing (unlike equilibrium processes) destroys the clustering and, after a finite amount of shear, restores one to states typical of the quasistatic sheared ensemble.  Moreover, we believe that the value of $\phi_J^{\rm QS}$ is independent of the specific details of the shearing dynamics, provided one is in the limit of overdamped particle motion; the $\phi_J^{\rm QS}$ we report here from quasistatic shearing, which involves energy minimization rather than a specific particle dynamics, agrees well with the value we find from a scaling analysis \cite{OlssonTeitel} of shear driven states at finite strain rates $\dot\gamma$ using Durian's mean-field bubble dynamics, Eq.~(\ref{em}).   

Thus, in our model, the athermal jamming of steady state shear driven systems occurs at a nontrivial (i.e. $\phi_J^{\rm QS}<\phi_{\rm max})$ unique point in the $(\phi,T,\dot\gamma)$ space that is independent of any further details of the shearing protocol.  
Our observation of critical scaling in such shear driven flow \cite{Olsson,OlssonTeitel}, leads us to conclude that $\phi_J^{\rm QS}$ locates a true {\it nonequilibrium} critical point in the $(\phi,T,\dot\gamma)$ phase space.  

The question of whether there can exist a sharp {\it equilibrium} glass transition in such simple models remains controversial.  Our results on jamming, reported here, make it interesting to speculate that, should such an equilibrium glass transition {\it not} exist, there may nevertheless be a sharp glass transition when one considers the behavior of shear driven steady states.

\section*{Acknowledgements}

This work was supported by Department of Energy Grant No. DE-FG02-06ER46298, Swedish Research Council Grant No. 2007-5234, a grant from the  Swedish National Infrastructure for Computing (SNIC) for computations at HPC2N and the University of Rochester Center for Research Computing.
We would like to thank L.~Berthier, R.~P.~Behringer, O.~Dauchot and M.~A.~Moore for helpful discussions.

\appendix
\section{}

In this section we provide several details concerning our simulation methods.  We first consider the stopping criterion $E/N<e_{\rm cut}=10^{-16}$ which we use to identify unjammed states.  Let $M_{\rm iter}$ be the total number of conjugate gradient iteration steps needed to achieve energy minimization of a particular configuration, according to the stopping criteria given in Section II.
In Fig.~\ref{f8} we show a scatter plot of the energy per particle $E/N$ vs the number of such conjugate gradient steps {\it per particle}, $M_{\rm iter}/N$, for all initial configurations of the RAND and QS ensembles, near the ensemble specific jamming density.
In Fig.~\ref{f8}a we show results for RAND for a system with $N=16384$ particles at a packing fraction $\phi=0.8415\approx\phi_J^{\rm RAND}$.  In Fig.~\ref{f8}b we show results for QS for a system with $N=4096$ at $\phi=0.8430\approx\phi_J^{\rm QS}$.  We see that the states clearly cluster into two groups:  those with $E/N=10^{-16}$, which correspond to the unjammed states for which our minimization has stopped upon reaching our lower cutoff treshhold $e_{\rm cut}$, and those with larger $E/N$, corresponding to the jammed states.  There are exceedingly few states, of negligible statistical weight, in the region connecting these two clusters; it is these states, with very small but finite energy, which take the longest to energy minimize.

\begin{figure}[h!]
\begin{center}
\includegraphics[width=3.3in]{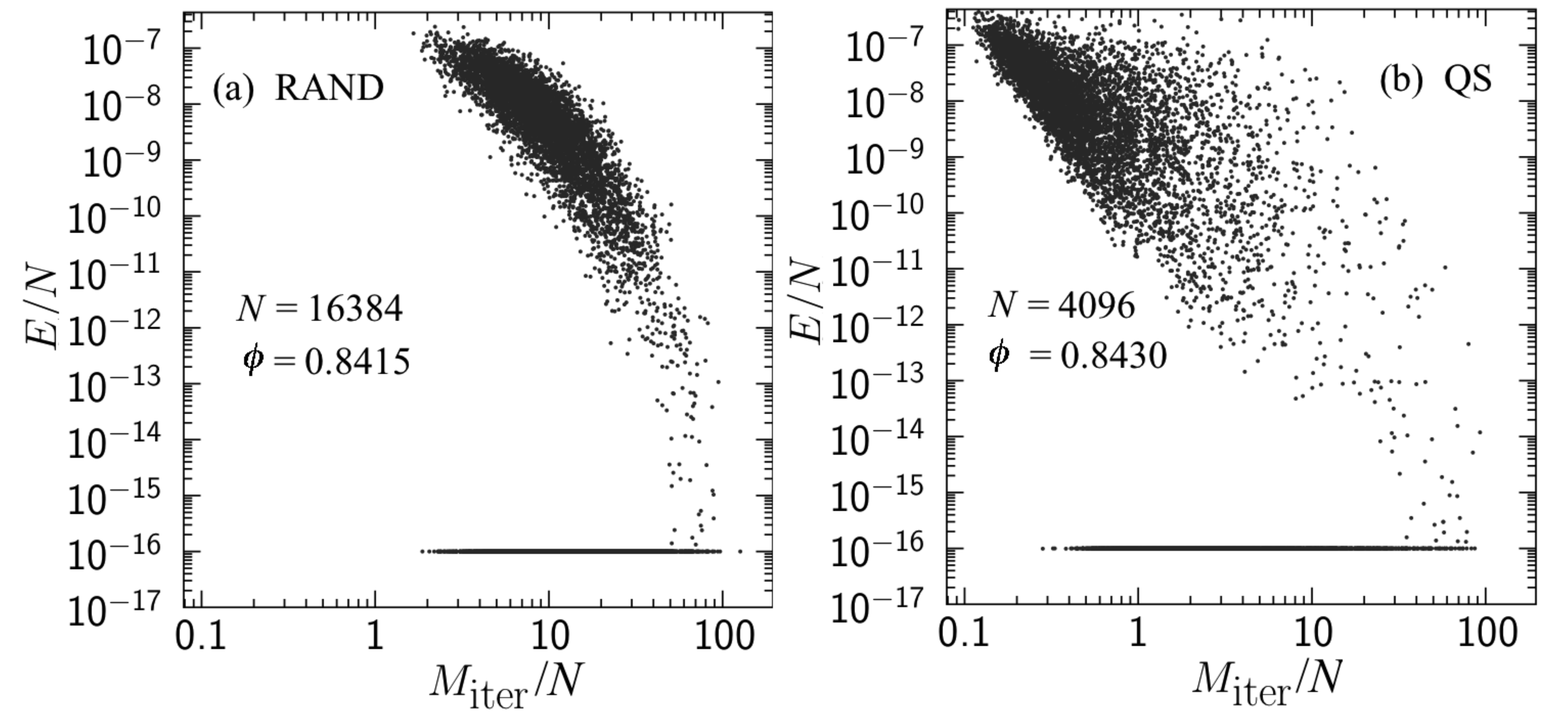}
\caption{Scatter plot of energy per particle $E/N$ vs the number of conjugate gradient iteration steps per particle $M_{\rm iter}/N$ needed to achieve minimization at $\phi\approx\phi_J$.  (a) is for RAND with $N=16384$ particles at $\phi=0.8415$; (b) is for QS with $N=4096$ particles at $\phi=0.8430$.
The horizontal line at $E/N=10^{-16}$ are the unjammed states, where the minimization has stopped upon reaching our cutoff threshold $e_{\rm cut}$.
}
\label{f8}
\end{center}
\end{figure}

In Fig.~\ref{f9} we plot the number of conjugate gradient iteration steps per particle $M_{\rm iter}/N$ needed to energy minimize the initial configuragtion, averaged over all initial configurations, vs the packing fraction $\phi$, for both RAND and QS, for several different system sizes $N$.  Note, since each initial state for RAND is completely random, while each initial state in QS starts as a small affine shear strain from a previously energy minimized state, the number of iterations needed for RAND is higher than that needed for QS.  As $N$ increases, $M_{\rm iter}(\phi,N)/N$ sharpens up to a peak located near $\phi_J$.
For RAND, we see that the peak value appears to be approaching a constant as $N$ increases.
For QS, we use $\Delta\gamma=10^{-4}$ for sizes $N\le 2048$, and for this strain increment, the peak number of iterations per particle also seems to be approaching a constant as $N$ increases.  For $N>2048$, however, we use $\Delta\gamma=10^{-5}$; each initial configuration is thus closer to its previous energy minimum from the prior strain step, and so takes fewer (about a factor of two) iteration steps to reach its new energy minimum than for the larger $\Delta\gamma$.  While decreasing $\Delta\gamma$ thus appears to make the energy minimization more efficient, as $\Delta\gamma$ decreases we obviously need to run longer to reach the same total strain $\gamma$.  Thus the net effect of decreasing the strain step from $\Delta\gamma=10^{-4}$ to $10^{-5}$ is about a factor 5 increase in computation.  It is interesting to note that for RAND, and also for QS for those sizes $N\le 2048$ where a constant $\Delta\gamma=10^{-4}$ is used, the curves of $M_{\rm iter}(\phi,N)$ for different $N$ all seem to intersect at roughly the same value $\phi^*\approx \phi_J$. If we interpret the total number of iterations needed to energy minimize, $M_{\rm iter}$, as a relaxation time $\tau$, then a finite size scaling analysis would suggest a divergent relaxation time $\tau\sim N\sim L^{z_{cg}}$ at $\phi_J$, with $z_{cg}\approx 2$.  We note, however, that the conjugate gradient ``dynamics" does not necessarily correspond to a real physical dynamics, so $z_{cg}$ need not equal the physical dynamic critical exponent $z$, such as one might find using the dynamics of Eq.~(\ref{em}).

\begin{figure}[h!]
\begin{center}
\includegraphics[width=3.3in]{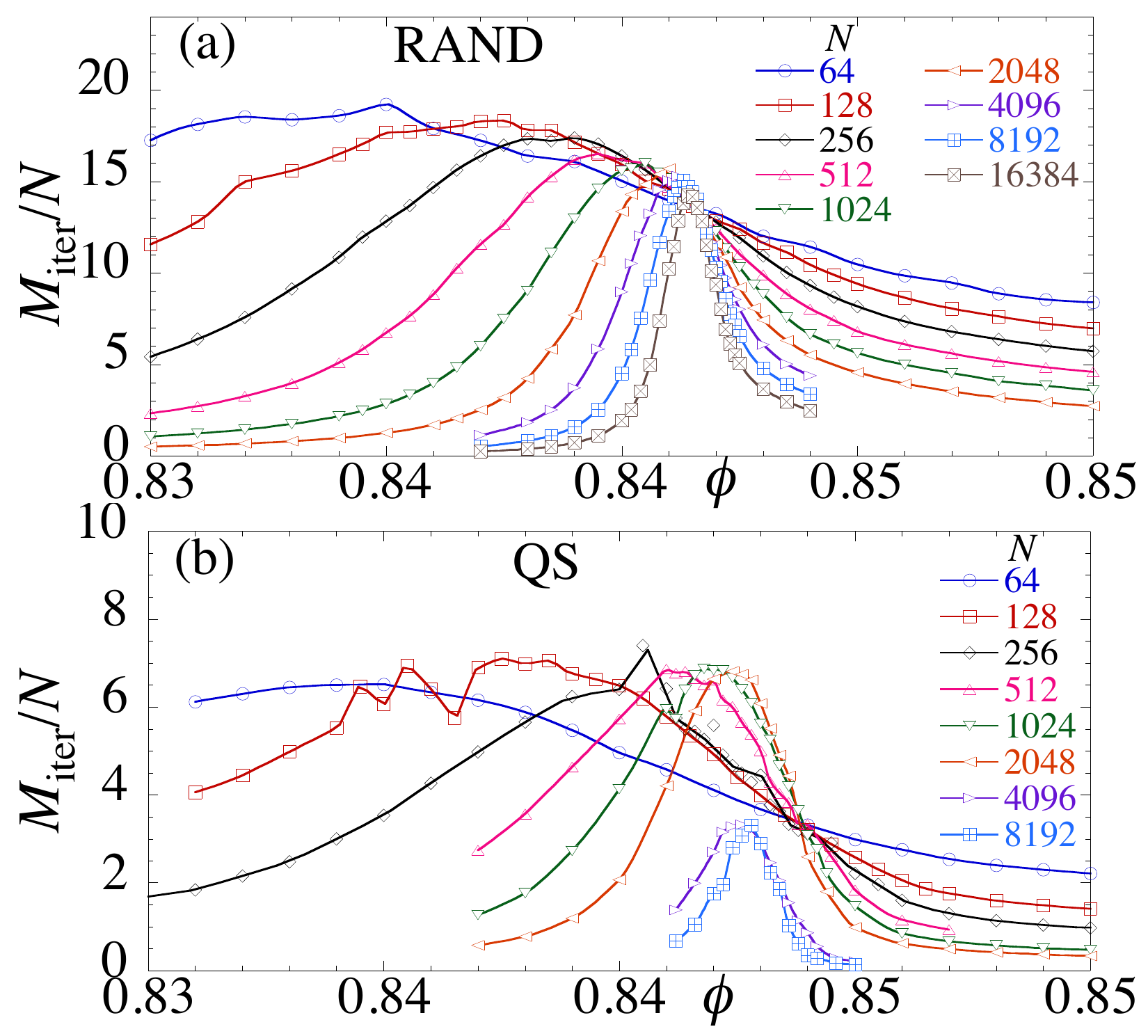}
\caption{(color online) Average number of conjugate gradient iteration steps per particle $M_{\rm iter}/N$ needed to energy minimize, vs packing fraction $\phi$.  (a) is for RAND; (b) is for QS.  Results are shown for a few different systems sizes with particle numbers $N$.}
\label{f9}
\end{center}
\end{figure}

Finally, in Fig.~\ref{f10} we consider the effect of the finite shear strain step $\Delta\gamma$ on our QS simulations.  In Fig.~\ref{f10}a we plot the fraction of jammed states $f$ vs $\Delta\gamma$ for the fixed value $\phi=0.8430\approx\phi_J^{\rm QS}$, for systems with different numbers of particles $N\le 4096$; in Fig.~\ref{f10}b we plot the corresponding energy per particle $E/N$ vs $\Delta\gamma$.  Note, at this value of $\phi$, the jammed fraction $f$ is in the range $0.5<f<0.8$, depending on system size $N$, hence we are adequately sampling the behavior in both jammed and unjammed states.  We see that for fixed $N$, both $f$ and $E/N$ decrease to a constant value (within the estimated statistical error \cite{eDg}) as $\Delta\gamma$ decreases.  The larger the value of $N$, the smaller $\Delta\gamma$ must be to reach the limiting constant value.
For the results reported in the main body of this work we have used $\Delta\gamma=10^{-4}$ for $N\le 2048$ and $\Delta\gamma=10^{-5}$ for $N\ge 4096$.  Comparison with Fig.~\ref{f10} shows that these values are small enough that we are in the correct quasistatic limit, except possibly for our largest system size $N=8192$; we have not been able to simulate $N=8192$ long enough with $\Delta\gamma<10^{-5}$ to verify that a smaller $\Delta\gamma$ is not needed for this case.

\begin{figure}[h!]
\begin{center}
\includegraphics[width=3.3in]{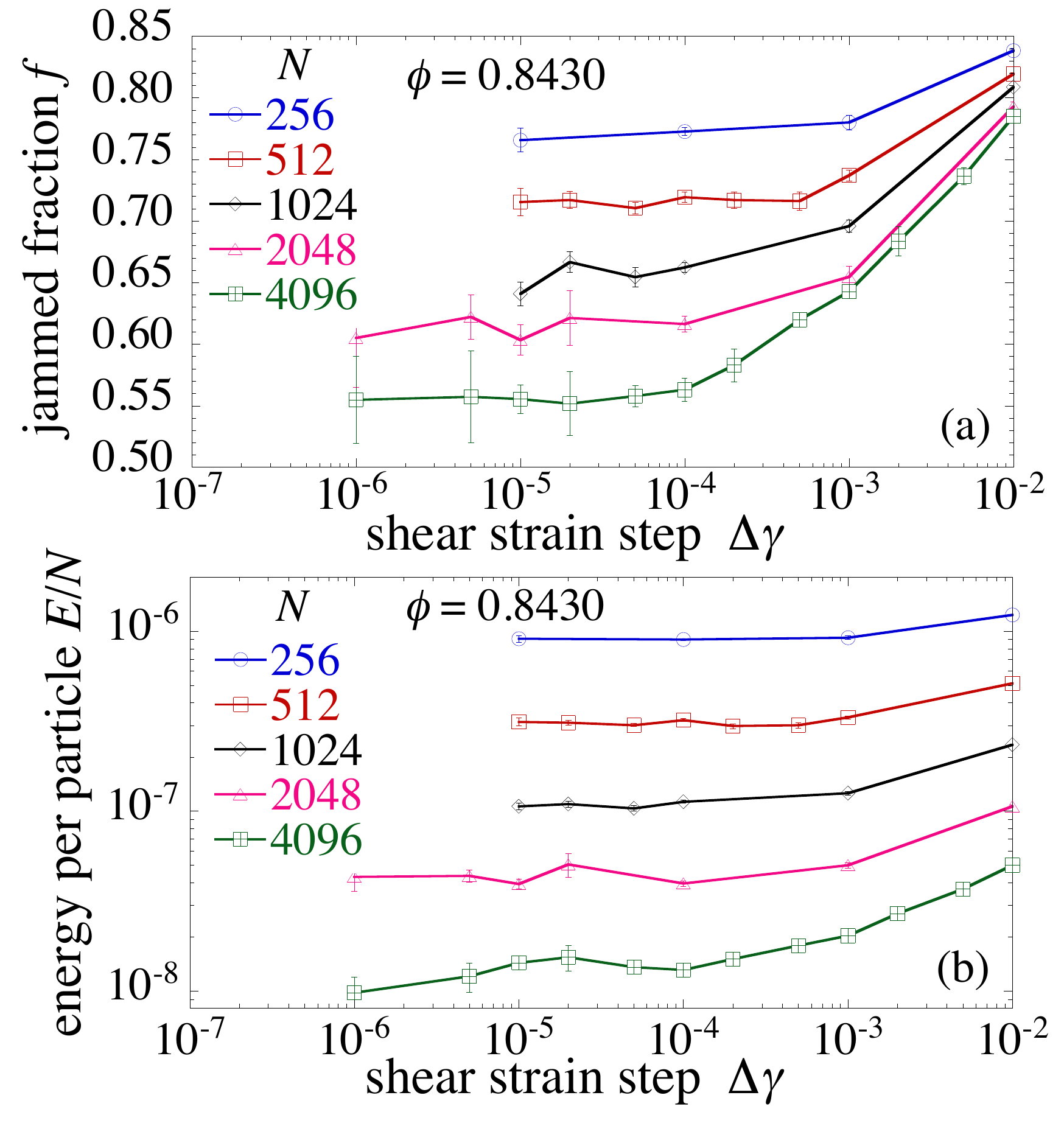}
\caption{(color online) (a) Dependence of jammed fraction $f$ and (b) energy per particle $E/N$ on shear strain step increment $\Delta\gamma$, within quasistatic shearing simulations, at fixed packing fraction $\phi=0.8430\approx \phi_J^{\rm QS}$.  Results are shown for several different system sizes with particle numbers $N$.}
\label{f10}
\end{center}
\end{figure}


\begin{thebibliography}{99}
\bibitem{note2}That $\phi_J^{\rm RAND}<\phi_J^{\rm QS}$ has also been noted in Ref.~\protect\cite{Heussinger2}, which we learned of as this work was being completed.
\bibitem{note1}A similar upper bound $\phi^*$ for jamming upon cooling  was found in 3D in Ref.~\protect\cite{Berthier1}.
\bibitem{note}Since Ref.~\protect\cite{Berthier3} is in 3D, it is possible that processes other than  the clustering we see in 2D may contribute to the dependence of $\phi_J$ on $\phi_{\rm init}$.
\bibitem{eDg}Note that our runs in Fig.~\ref{f10} are in general not as long as those used to obtain the data shown in Fig.~\ref{f1}, hence our error bars here are generally larger.
\end{thebibliography}
\end{document}